# Designing optoelectronic properties by on-surface synthesis: formation and electronic structure of an iron-terpyridine macromolecular complex

*Agustin Schiffrin*[||,~,†,§,‡], *Martina Capsoni*[||,‡], *Gelareh Farahi*[||], *Chen-Guang Wang*[^], *Cornelius Krull*[†], *Marina Castelli*[†], *Tanya S. Roussy*[||], *Katherine A. Cochrane*[#], *Yuefeng Yin*[§,±,†], *Nikhil Medhekar*[§,±], *Adam Q. Shaw*[||], *Wei Ji*[^,*] *and Sarah A. Burke*[||,#,~,*]

[||]Department of Physics and Astronomy, University of British Columbia, Vancouver British Columbia, Canada, V6T 1Z1

[~]Quantum Matter Institute, University of British Columbia, Vancouver British Columbia, Canada, V6T 1Z4

[†]School of Physics & Astronomy, Monash University, Clayton, Victoria 3800, Australia

[§]ARC Centre of Excellence in Future Low-Energy Electronics Technologies, Monash University, Clayton, Victoria 3800, Australia

[#]Department of Chemistry, University of British Columbia, Vancouver British Columbia, Canada, V6T 1Z1

[^]Department of Physics and Beijing Key Laboratory of Optoelectronic Functional Materials & Micro-nano Devices, Renmin University of China, Beijing 100872, People's Republic of China

[±]Department of Materials Science and Engineering, Monash University, Clayton, Victoria 3800, Australia

ABSTRACT: Supramolecular chemistry protocols applied on surfaces offer compelling avenues for atomic scale control over organic-inorganic interface structures. In this approach, adsorbate-surface interactions and two-dimensional confinement can lead to morphologies and properties that differ dramatically from those achieved via conventional synthetic approaches. Here, we describe the bottom-up, on-surface synthesis of one-dimensional coordination nanostructures based on an iron (Fe)-terpyridine (tpy) interaction borrowed from functional metal-organic complexes used in photovoltaic and catalytic applications. Thermally activated diffusion of sequentially deposited ligands and metal atoms, and intra-ligand conformational changes, lead to Fe-tpy coordination and formation of these nanochains. Low-temperature Scanning Tunneling Microscopy and Density Functional Theory were used to elucidate the atomic-scale morphology of the system, providing evidence of a linear tri-Fe linkage between facing, coplanar tpy groups. Scanning Tunneling Spectroscopy reveals highest occupied orbitals with dominant contributions from states located at the Fe node, and ligand states that mostly contribute to the lowest unoccupied orbitals. This electronic structure yields potential for hosting photo-induced metal-to-ligand charge transfer in the visible/near-infrared. The formation of this unusual tpy/tri-Fe/tpy coordination motif has not been observed for wet chemistry synthesis methods, and is mediated by the bottom-up on-surface approach used here.

A crucial challenge for (opto)electronics technologies relying on organic molecules and polymers remains control over solid-state structure, particularly at interfaces.[1-2] Bottom-up supramolecular self-assembly techniques offer exquisite control over nanoscale structure and properties, allowing for synthesizing mesoscopic structures from carefully designed molecular tectons, with atomic-scale precision and without intervention[3]. When applied on a surface, these approaches have yielded a wide range of low-dimensional supramolecular systems with high fidelity,[4-5] yet few have demonstrated the emergence of deliberately sought physicochemical functionalities.[6-8] The robust bonding[9-10] and remarkable range of physical and chemical properties found in coordination complexes, tuned via choice of the metal center and ligand environment, offers a promising avenue for engineering both structure and functionality via self-assembly.[11-12] In particular, polypyridyl-based metal-organic complexes and coordination polymers exhibit an extensive range of opto-electronic, spin, and chemical properties that can be exploited in technological applications, such as photovoltaics,[13] catalysis,[14-15] molecular electronics,[16-17] molecular magnetism,[18] and biomedicine.[19-20] Complexes consisting of group 8 transition metals – iron, ruthenium, and osmium – coordinated with bis- and terpyridine (tpy) ligands have received special attention due to their photophysical properties, including optical transitions (from the ultraviolet to near infrared) related to intramolecular photo-induced metal-to-ligand charge transfer (MLCT). These properties can give rise to efficient photo-induced charge separation needed for photovoltaics,[21-22] and photocatalysis.[23] The robustness and well-defined morphology of the metal-polypyridyl coordination motif further allows for the design of hierarchical metallo-supramolecular structures[24-26] with a similarly rich array of physical properties, while offering additional avenues of control over the structure and environment by tuning both intra- and intermolecular interactions through ligand modification.[27-28]

Extending the functional motif of these coordination complexes to surface-bound protocols of supramolecular chemistry[4, 29-30] offers practical advantages in terms of both synthesis and nanoscale control for device design, e.g., in photovoltaics and heterogeneous catalysis, where active materials are in contact with a solid substrate. However, the surface introduces both challenges and opportunities to the robust design of bottom-up nanostructures: interactions with the substrate may alter the intended intrinsic properties, potentially allowing for the formation of unique structures with coordination symmetries, metal oxidation states, and polyatomic metal centers[9, 31-32] that are markedly different from those achieved via wet-chemistry-based synthetic methods [i.e., where precursors are not confined to two dimensions (2D)]. Such distinct structures will also manifest distinct electronic properties through changes in the underlying electronic states.

Here, we describe the formation, morphology and electronic structure of linear self-assembled metal-organic nanostructures via thermally activated coordination of bis-terpyridine (bis-tpy) based ligands (terpyridine-phenyl-phenyl-terpyridine; TPPT) with iron (Fe) adatoms on a Ag(111) surface. These surface supported macromolecular complexes are synthesized through sequential and independent deposition of the ligand and of the transition metal, providing a flexible methodology to tune the electronic properties of the resulting structures. Scanning Tunneling Microscopy (STM) and Spectroscopy (STS) performed in ultrahigh vacuum (UHV) at low-temperature, together with Density Functional Theory (DFT), were used to determine the atomic-scale morphology and electronic states of the metal-organic chains. Our STM data, supported by DFT calculations, hint towards an unusual trinuclear Fe coordination linkage between coplanar tpy groups of facing flat TPPT ligands, enabled by the bottom-up on-surface synthetic approach, very different from related wet-chemistry-synthesized macromolecular complexes. Notably, our STS measurements reveal an electronic structure indicative of an MLCT absorption band, reminiscent of optical excitations in the visible/near-infrared (VIS/NIR) of octahedral Fe(tpy)$_2$ complexes in solution.[33-34] These surface-bound macromolecular structures offer new opportunities for controlling optoelectronic and catalytic function.

RESULTS

Figure 1(a) shows a STM image of TPPT molecules on Ag(111), before the deposition of Fe. The molecule – imaged as a "dog-bone" – adsorbs on the surface with a planar geometry, similar to other tpy containing molecules.[35] Details on the nanoscale adsorption morphology and electronic structure of TPPT on Ag(111) and can be found elsewhere.[36] The molecules can bind to each other laterally, via attractive non-covalent interactions between peripheral pyr groups, forming staggered rows or zigzag patterns[36] (Figure 1a).

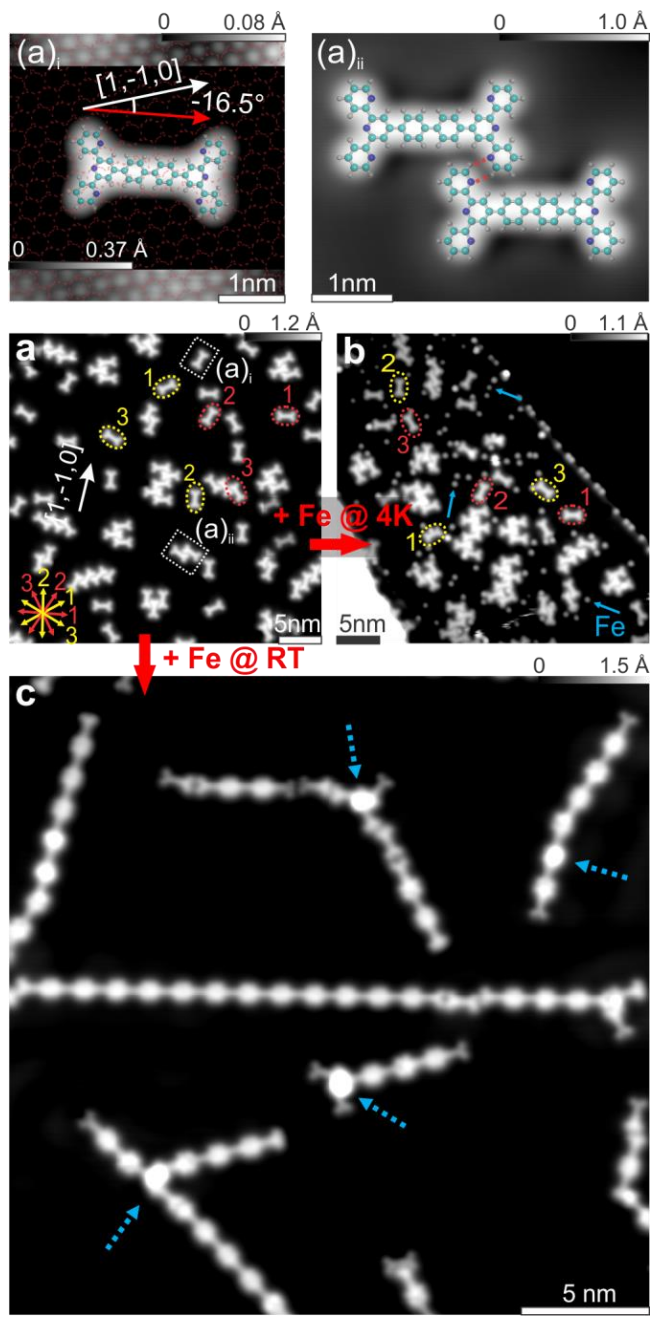

**Figure 1.** Low-temperature STM imaging of metal-organic nanochain formation. (a) Terpyridine-phenyl-phenyl-terpyridine (TPPT) molecule deposited on Ag(111) at room temperature ($V_b$ = 200 mV, $I_t$ = 50 pA). Molecules adsorb at a ~±16.5° (red and yellow dashed ellipses) angle with respect to the <1, 1, 0> crystallographic directions (white arrow), resulting in six energetically equivalent orientations (labels 1, 2 and 3). (a)$_i$ Single TPPT molecule

(molecule: $V_b$ = -200 mV, $I_t$ = 10 pA; Ag(111): $V_b$ = -10 mV, $I_t$ = 300 nA ) and (a)$_{ii}$ TPPT dimer formed via hydrogen bonds between peripheral pyridine groups ($V_b$ = -100 mV, $I_t$ = 50 pA). Blue: nitrogen, cyan: carbon; white: hydrogen. (b) Fe adatoms (solid turquoise arrows) deposited on TPPT/Ag(111) at ~4.3 K ($V_b$ = 500 mV, $I_t$ = 10 pA). (c) Fe deposited at room temperature on TPPT/Ag(111) ($V_b$ = 20 mV, $I_t$ = 25 pA), inducing the formation of metal-organic chains. Dashed turquoise arrows show Fe clusters.

Figure 1b shows an STM topograph after Fe deposition onto the TPPT/Ag(111) system held at 4.3 K (see Methods). The Fe adatoms (bright protrusions marked with solid turquoise arrows in Figure 1b) are scattered randomly on the surface and do not perturb the TPPT motifs observed prior to their addition (Figure 1a). After annealing at 373 K the Fe/TPPT/Ag(111) system prepared at 4.3 K, or following Fe deposition onto TPPT/Ag(111) at RT (Figure 1c), the atomic and molecular species rearrange into linear nanostructures. In these nanochains, TPPTs no longer bind laterally to each other, as in Figure 1a, but in a head-to-head configuration. Some of the Fe atoms also bind to step edges and aggregate into clusters[37] (dashed turquoise arrows in Figure 1c). At 4.3 K, transition metal adatoms[38-39] and nitrogen-containing aromatic molecules[40-41] are immobile and do not diffuse on close-packed noble metal surfaces. At RT and above, both molecules and metal adatoms can diffuse, with diffusion rates that at RT are at least an order of magnitude larger for group 8 transition metal atoms[38-39] than for conjugated organic molecules;[40] thermal activation is necessary to enable mobility of both components. Importantly, an increase in average chain length can be achieved by annealing at higher temperature (e.g. 373 K) or by increasing the annealing time (e.g., on the order of several minutes at RT). This increase is accompanied by a decrease of isolated metalated molecules (i.e., not incorporated in chains, and with one or two tpy sites coordinated with Fe). This indicates that isolated Fe-coordinated TPPT serve as building blocks for the self-assembly of the coordination complexes.

To understand the first step in forming Fe-TPPT chains, we investigated the configuration of isolated TPPT molecules after Fe deposition, in which one or both tpy groups have interacted with Fe adatoms. Figure 2b shows an STM image of an isolated TPPT molecule where the left-hand tpy is identical to that of an isolated TPPT before[36] Fe deposition (see Figure 1a). The right-hand tpy appears brighter, with a central protrusion. The latter was not observed prior to Fe deposition. This is the result of at least one Fe adatom that has interacted with the tpy group. Since all observed tpy groups of TPPT show one of these two imaging characteristics, we conclude that the termination consists of either zero or one Fe atom, but not more, as this would appear distinct in STM imaging. This is confirmed by DFT simulations (see Figure 2e and Methods).

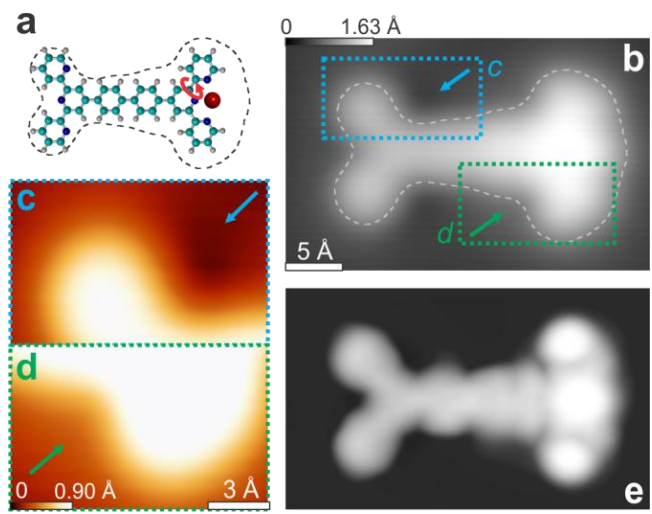

**Figure 2.** (a) Chemical structure (red: iron) and (b) STM image of single TPPT with right tpy group coordinated to one Fe adatom ($V_b$ = -200 mV, $I_t$ = 1 nA). Coordination mediated by the rotation [red arrow in (a)] of peripheral pyr groups, which in the case of non-metalated tpy [left

tpy in (a)] point toward the outside of the molecule. Black dashed curve in (a) indicates molecular contour on STM image in (b). (c), (d) Zoomed in details of regions *c* and *d* of STM image in panel (b). Non-metalated tpy shows a depression (turquoise arrow) next to the peripheral pyr, not observed (green arrow) for metalated tpy (d). (e) DFT-simulated STM image ($V_b$ = -200 mV) of TPPT molecule with right tpy group coordinated to one Fe adatom, according to structure in (a).

Closer examination of the negative bias STM images reveals small apparent depressions in the substrate (turquoise arrows in Figures 2b and c) adjacent to the outer pyr rings. These depressions, indicative of a lowered electronic density of the Ag substrate, can be explained by repulsion of the surface conduction electrons by the outward pointing distal pyr groups. Conversely, the absence of these depressions for Fe-coordinated tpy groups (green arrows in Figure 2b, d) indicates a rotation of the distal pyridines toward the metal adatom, allowing them to fully participate in the metal-ligand coordination, again consistent with the DFT results (Figure 2e).

The molecular configuration with all three pyr rings of a metalated tpy group oriented with the N atoms towards the Fe has two implications for the adsorption geometries. The coordination with Fe, involving all three pyr's (which previously interacted significantly with the substrate), reduces the interaction of the molecule with the surface. This is consistent with the observed loss of preferential orientation of the Fe-TPPT chains with respect to the crystalline axes of the substrate (Figure 1c). Also, the zig-zag and staggered intermolecular configurations, observed for pristine ligands[36] (Figure 1a) and characterized by lateral noncovalent interlocking between outward-pointing tpy groups of adjacent molecules, were not seen for tpy groups coordinated with Fe. Hence, thermal annealing of the Fe/TPPT system activates a configuration change of the tpy, enabling Fe-tpy coordination and altering intermolecular and molecule-substrate interactions.

Bias-dependent STM imaging and normalized $(dI/dV)/(I/V)$ STS[42] provide further insight into the morphology and electronic properties of the metalated molecule (Figure 3). Figure 3a shows an occupied-state STM image of a single TPPT, where each tpy group is coordinated with a single Fe atom (indicated by white arrows), for a sample bias voltage of -0.5 V. Here, each tpy group is imaged identically to the right metalated tpy in Figure 2b, with a significant contribution to the imaging from the center of the tpy, where the Fe atom is located. Figure 3b shows an unoccupied-state STM image of the same metalated molecule, at +0.1 V. At this positive bias, the metalated tpy groups are v-shaped, similar to the non-metalated tpy (see left tpy in Figure 2b); the areas where the Fe atoms are located contribute negligibly to the imaging. This is a strong indication that occupied Fe electronic states near the Fermi level contribute significantly to the highest occupied molecular orbitals (HOMOs) of the system, whereas empty Fe states have negligible contributions to the lowest unoccupied levels (LUMOs). This is emphasized in Figure 3c, corresponding to the substraction of these images; bright regions indicate areas that contribute more to the HOMOs than to the LUMOs, with a clear bright protrusion at the center of the tpy group, where the Fe atom is located.

Figure 3d displays $(dI/dV)/(I/V)$ spectra taken at the center of the single pristine (blue curve in Figure 3d) and doubly metalated (red) TPPT. The latter shows a clear tunneling resonance at ~+1.5 V, given by contributions of ligand-related unoccupied states to the density of states (DOS). Notably, this feature is ~+0.3 V above the tunneling resonance observed for pristine TPPT.[36]

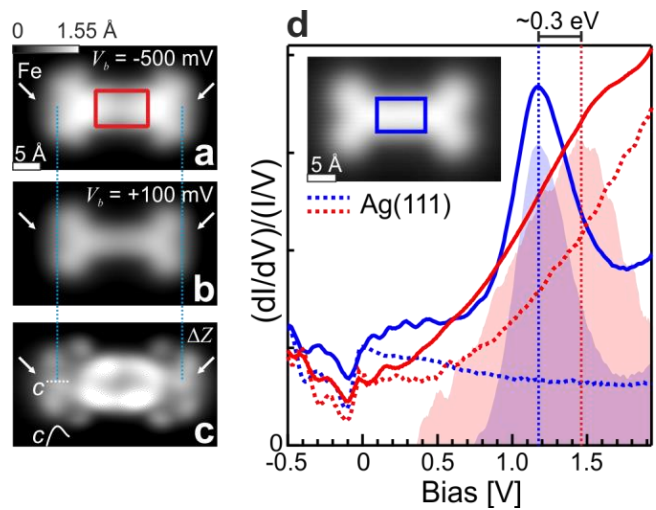

**Figure 3.** (a) Negative (-500 mV) and (b) positive (+100 mV) bias STM images of single TPPT molecule with both tpy groups metalated with Fe adatoms ($I_t$ =200 pA). (c) Difference STM topographic map resulting from subtraction of (a) with (b). White arrows indicate the position of single Fe adatoms. Height profile *c* along molecular axis over metalated tpy shows a protrusion due to Fe atom. Vertical dashed turquoise lines show positions of Fe atoms. (d) Average (d$I$/d$V$)/($I$/$V$) spectra at center of single non-metalated (blue) and Fe-coordinated (at both tpy groups; red) TPPT molecule. Dashed curves: spectra on bare Ag(111). Background filled curves (rescaled and offset for clarity): difference spectra resulting from the substraction of solid blue and red curves with dashed curves, to emphasize tunneling resonances. Within the bias voltage window considered, we did not observe clear tunneling resonances associated with occupied ligand states. Inset: STM image of single, non-metalated TPPT ($V_b$ = -2.5 V, $I_t$ = 50 pA). Rectangles indicate areas where spectra were taken.

We now focus on structural details of the metal-organic nanochains. Similar to Figure 3, Figures 4a, b show high-resolution STM images of a chain composed of five TPPT molecules at different biases. The chains have a periodicity of (2.3 ± 0.2) nm. At a bias voltage of -0.5 V, STM imaging shows taller features at the Fe coordination centers. At +0.1 V, these regions are imaged as slight depressions compared to the surrounding ligand. As for the metalated TPPT in Figure 3, this indicates a significant contribution of the Fe electronic states to the occupied DOS in this region, whereas their contribution to the unoccupied DOS is limited. This is emphasized in the subtraction of these two topographic maps (Figure 4c). At each coordination center, we observe two clear protrusions separated by a distance of (5 ± 1) Å, that we can fit by two Gaussian curves (white profiles in Figure 4c).

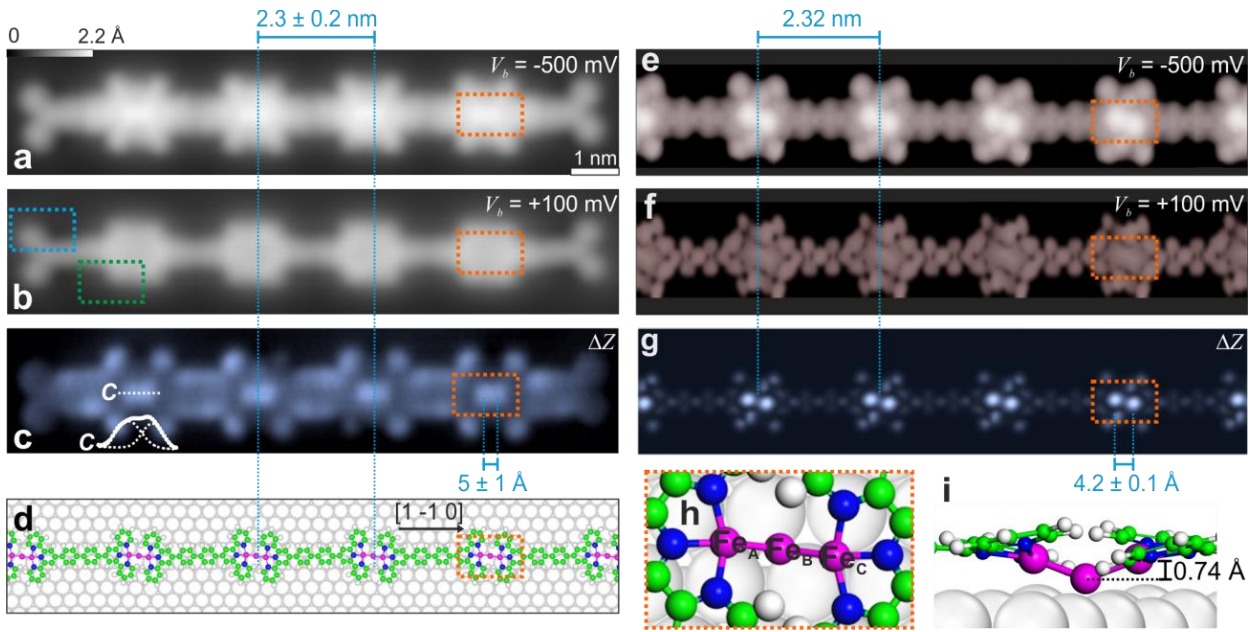

**Figure 4.** (a) Negative (-500 mV) and (b) positive (+100 mV) bias high-resolution STM images of a 5-molecule Fe-TPPT nanochain. ($I_t$ = 1 nA). (c) Difference STM topographic map resulting from subtraction of (a) with (b). Height profile $c$ along coordination center shows two protrusions fitted with Gaussians. (d) Model of the DFT-optimized metal-organic nanochain. (e) Negative and (f) positive bias DFT-simulated topographic STM images of the optimized Fe-TPPT/Ag(111) configuration. (g) DFT-simulated difference STM topographic map resulting from subtraction of (e) and (f). (h), (i): Top and side view of the DFT-optimized coordination center.

In order to better understand the structure of these chains, we modeled Fe-TPPT chains by DFT (see Methods) using periodic boundary conditions along the [1, -1, 0] axis of Ag(111) and a spacing of at least 8.7 Å between two neighboring chains along the [1 1 -2] direction. Coordination linkages consisting of one, two or three Fe adatoms were considered. The structure with three linearly arranged Fe atoms between the flat-lying tpy groups of the TPPT ligands (Figures 4d, h, i) was found to be most energetically favorable by 0.57 eV, with a periodicity of 2.32 nm in excellent agreement with the experiment. A coordination configuration involving one or two adatoms is hindered by steric repulsion between the flat-lying tpy's. STM images simulated from the DFT model [Figures 4e, f; details in Supplementary Information (SI)] reproduce the experimental data well (Figures 4a, b). Importantly, in the topographic map subtraction in Figure 4g, the 3-Fe coordination configuration appears as two bright protrusions separated by (4.2 ± 0.1) Å, given by the two tpy-bound Fe atoms ($Fe_A$ and $Fe_C$ in Figures 4h, i) being 0.74 Å higher than the central Fe ($Fe_B$), in good agreement with the experiment (Figure 4c). This agreement between simulated and experimental STM images supports the claim that the inter-ligand binding is mediated by a tpy-triiron-tpy coordination scheme. Further support for this linear tri-iron structure is provided in the SI. While similar tri-metallic clusters have been observed for copper[32] (without the participation of the outer pyr's of the tpy group), this is a highly unusual and unexpected structure for Fe, mediated by the 2D confinement on the surface and facilitated by the UHV environment.

We explain the chain formation mechanism as follows. Upon deposition of atoms and molecules, thermal activation enables adsorbate diffusion and rotation of the TPPT distal pyr rings, allowing the tpy groups to coordinate with the Fe adatoms. The noble metal surface is crucial since it confines the system to 2D, enables efficient diffusion and does not alter the molecular chemistry.

Chain nucleation occurs when two metalated TPPT's *capture* and *lock* an additional Fe adatom. It is important to note that the observed metal-organic coordination cannot be explained by bridging residual gas ligands (e.g., CO, $H_2O$) or adatoms from the Ag substrate (see SI).

We now turn to the electronic structure of this coordination complex. Figure 5a shows d$I$/d$V$/($I$/$V$) spectra[42] for the Fe center (orange) and the centre of the TPPT ligand (red) in the metal-organic nano-chain, compared with the pristine non-metalated molecule (blue). Similarly to the single, double metalated TPPT molecule in Figure 3, a ligand in a nanochain [see top-right STM topograph in Figure 5(a)] exhibits a clear tunneling resonance at a bias voltage of ~ +1.5V, that is, ~ +0.3V above the strong tunneling resonance associated with empty states located at the centre of the pristine non-metalated TPPT; a single TPPT molecule with both its *tpy* groups coordinated to a Fe adatom is similar (electronically) to a TPPT ligand within a metal-organic nanochain. Within the considered positive bias range (empty states), the differential conductance acquired on a Fe coordination center [orange curve in Figure 5(a)] does not show any clear tunneling resonance, and is dominated by the exponential tunneling transmission function. However, a clear feature is observed for the Fe node at ~ -0.09 V. At this negative bias voltage, no feature was observed at the center of the pristine, non-metalated TPPT[36], or at the centre of the single Fe-TPPT-Fe in Figure 3d. This feature is emphasized by the background grey filled curve in the inset of Figure 5a, corresponding to the ratio between the differential conductance at the Fe center and that at the ligand centre. It is important to note that this negative bias (d$I$/d$V$)/($I$/$V$) feature associated with the Fe node was also clearly observed in non-normalized d$I$/d$V$ spectra and local density of states spectra (retrieved according to Passoni *et al.*[43]; see Figure S4 in SI); it is an intrinsic spectroscopic signature of occupied electronic states near Fermi located at the Fe coordination node.

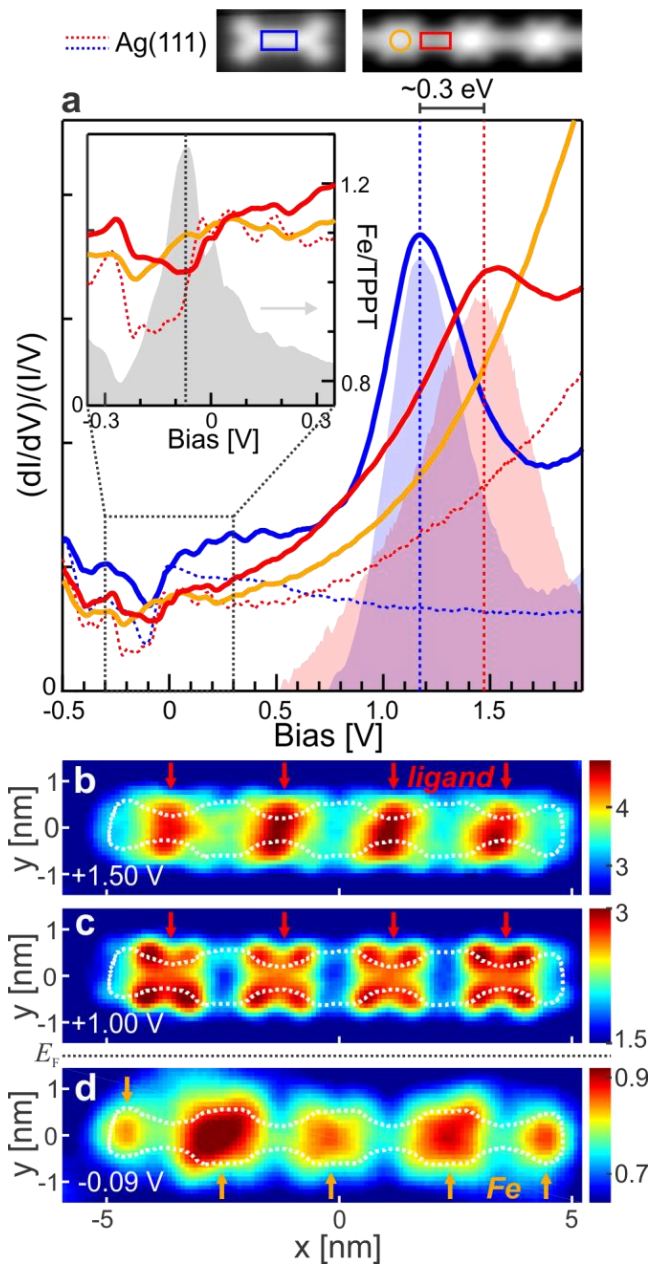

**Figure 5** (a) Average (d$I$/d$V$)/($I$/$V$) spectra at center of single non-metalated (blue) TPPT molecule (setpoint: $V_b$ = -2.5 V, $I_t$ = 50pA), of ligand in Fe-TPPT chain (red) and of Fe coordination node (orange) ($V_b$ = -2.5 V, $I_t$ = 2.5 nA). Dashed curves: spectra taken on bare Ag(111). Above: STM images of single non-metalated TPPT molecule (left; $V_b$ = -2.5 V, $I_t$ =

50pA) and Fe-TPPT nanochain (right; $V_b$ = -2.5 V, $I_t$ = 2.5 nA). Rectangles and circle indicate areas where spectra were taken. Background filled curves (rescaled and offset for clarity): difference spectra resulting from the substraction of solid (blue, orange, red) curves with corresponding dashed curves. Inset: detail of (d$I$/d$V$)/($I$/$V$) spectra near the Fermi level; background grey filled curve corresponds to the ratio between spectra taken on the Fe node (orange) and ligand (red) in Fe-TPPT nanochain. (b) – (d) (d$I$/d$V$)/($I$/$V$) maps of a Fe-TPPT nanochain at $V_b$ = -0.09, 1.00 and 1.50 V. Red (orange) arrows indicate positions of ligands (Fe coordination nodes) in (b) and (c) [(d), respectively].

DISCUSSION

Our STS data unequivocally show that the HOMOs of the metal-organic system have a dominant contribution from states located at the Fe coordination node (Figure 5d), and LUMOs with dominant contributions from ligand-related states (Figures 5b, c). These electronic properties are similar to those of complexes consisting of polypyridyl ligands coordinated in a octahedral geometry with a single Fe(II) atom, where Fe-dominated HOMOs and ligand-dominated LUMOs give rise to an visible light absorption band at photon energies $\hbar\omega$ > ~2 eV (wavelengths $\lambda$ < ~600 nm).[22, 33-34, 44-46] Interestingly, the electronic properties of our flat, tri-iron complex are similar to these orthogonal, single Fe ones, despite the different coordination morphologies. In our case, the energy difference between the strong ligand-associated tunneling resonance at ~1.5 eV (Figures 5a, b) and that related to the Fe node (~ -0.09 eV; Figures 5a, d) would potentially correspond to an optical transition in the VIS/NIR (at $\hbar\omega$ < ~1.59 eV, $\lambda$ > ~780 nm, given that the energy difference in the electronic tunneling spectrum in Figure 5a would typically be larger, by the exciton binding energy, than the energy involved in the associated photoexcitation). Here, the polarizability of the underlying noble metal substrate might arguably reduce this corresponding electronic energy difference by screening the tunneling electrons and decreasing (increasing) the ionization potential (electron affinity, respectively); less polarizable substrates might result in larger electronic energy differences.

The likelihood of a potential optical transition associated with a photo-induced Fe-to-TPPT electron transfer would scale as $\propto \left|\int \varphi_i^*(\mathbf{r})(\nabla \varphi_f(\mathbf{r})) \, d\mathbf{r}\right|^2$, according to Fermi's golden rule,[47] with $\varphi_i(\mathbf{r})$ being the molecular orbital associated with the high $(dI/dV)/(I/V)$ at the Fe node (Figure 5d) and $\varphi_f(\mathbf{r})$ that at the ligand (Figures 5c, d). Here, the combination of spatial separation between the *centres of mass* of empty (Figures 5b, c) and occupied (Figure 5d) states, and at the same time significant spatial overlap between such states, can potentially give rise to efficient photo-absorption and photo-induced electron-hole separation in the VIS/NIR, useful for photovoltaics and photocatalysis.

The positive energy shift (~ +0.3 eV) of the ligand-associated $(dI/dV)/(I/V)$ peak (~ +1.2 eV for the pristine TPPT; ~ +1.5 eV for TPPT in chain; blue and red curves in Figure 5a) that accompanies the Fe-TPPT coordination can be explained by a decrease of the electron affinity (EA) of the molecule upon metal-ligand coordination. This decrease of EA is similar to that observed for an isolated, metalated TPPT which is not part of a chain (Figure 3d). This effect can be due to two phenomena: (i) reduction of interaction and electronic coupling between TPPT and Ag(111) upon Fe-tpy coordination, resulting in less screening of tunneling charge by the polarizable metal substrate [resulting in a $(dI/dV)/(I/V)$ peak energy resembling that of TPPT on an electronically decoupling NaCl layer on[36] Ag(111)]; (ii) Fe-to-TPPT electron transfer upon Fe-TPPT coordination. A coordination-induced reduction of the TPPT-surface electronic interaction would arguably result in a narrower, sharper (d$I$/d$V$)/($I$/$V$) peak [on NaCl, the full width at half maximum (FWHM) of the (d$I$/d$V$)/($I$/$V$) peak related to the centre[36] of TPPT is ~0.2 eV]. This was not observed in our experiments (FWHM of background red filled curve in Figure 5a is ~0.65 eV). We can hence conclude that the observed decrease of ligand EA is most likely due to a coordination-induced metal-to-ligand electron transfer. This provides a further analogy and is consistent with the coordination-induced Fe-to-ligand electron transfer observed for single-Fe polypyidyl complexes with a octahedral coordination motif, where the Fe atom is in a +2 oxidation state.[45] A quantitative determination of the electronic configuration of each Fe atom in our metal-organic nanochains is beyond the scope of this study and requires further investigations.

The (d$I$/d$V$)/($I$/$V$) spectroscopic signature at ~ -0.09 eV associated with the Fe node (Figures 5a, d) lies at an energy slightly smaller than that of the Ag(111) Shockley surface state (SS) onset. This feature could be explained by a bound state located at the Fe node resulting from the coupling between an Fe energy level and the SS band.[48-50] However, a spectroscopic feature due to SS-localization would be significantly different (see SI). In our case, the coplanar tpy's form a cavity that scatters the SS, minimizing the Fe-SS interaction and the likehood of such coupling. Importantly, the spectroscopic signature of such SS-related bound state – which is claimed to be independent of the adsorbate[49] – was not observed for similar Cu-based metal-organic complexes on Cu(111) reported in a previous study.[32] In this previous study,[32] no HOMOs with dominant contributions from Cu adatom states were reported. We therefore ascribe our observed ~ -0.09 eV (d$I$/d$V$)/($I$/$V$) feature to electronic properties intrinsic to the Fe-tpy node.

CONCLUSIONS

In summary, we have shown the on-surface, bottom-up synthesis of metal-organic nano-structures based on the coordination of group 8 transition metal atoms with tpy compounds. The growth process is thermally driven, and enabled by the on-surface approach, which provides 2D confinement and stabilizes the macromolecular complexes. Our experimental observations, supported by DFT, point to a coordination motif consisting of a linear Fe trimer bonded to the coplanar tpy groups of the ligand, and cannot be achieved via conventional wet chemistry synthesis. The electronic structure of the system, including highest occupied (lowest unoccupied) states predominantly located at the transition metal centre (ligand, respectively), can potentially give rise to photo-induced metal-to-ligand charge transfer in the VIS/NIR, resulting in photophysical properties useful for photovoltaics and light-to-energy conversion technologies, in which atomic-scale structural control can lead to performance enhancement. Furthermore, the on-surface nanostructuring of well-defined group 8 transition metal clusters provides an interesting platform for site-isolated, heterogeneous catalysis with multi-metallic centers. Our approach offers new pathways to investigate and tailor the optoelectronic properties and reactivity of metal-organic nanostructures from the bottom-up.

METHODS

The metal-organic nanostructures were synthesized in UHV by sequential deposition of TPPT molecules (HetCat Switzerland) and Fe atoms from the gas phase onto a clean Ag(111) surface. The latter was prepared by repeated cycles of $Ar^+$ sputtering and annealing at 790 K. TPPT was sublimed at 550 K onto the Ag substrate held at RT, at a deposition rate of ~4 x $10^{-4}$ molecules/($nm^2$ s). The Fe atoms were subsequently evaporated onto the surface held either at 4.3 K (Fig. 1b) or RT (Fig. 1c). All STM and STS measurements were performed at 4.3 K with an Ag-terminated Pt/Ir tip. All topographic images were acquired in constant-current mode. STS measurements were obtained by measuring the tunneling current as a function of tip-sample bias voltage and tip position on the sample, with the tip height stabilized according to a constant-current set point at each location. The normalized numerical derivative[42] d$I$/d$V$/($I$/$V$) curves were computed as an approximation of the local density of states. The sample bias voltage is reported throughout the text. The base pressure was below 2 x $10^{-9}$ mbar during molecular deposition, and below 1 x $10^{-10}$ mbar during Fe evaporation and STM and STS measurements.

DFT calculations were carried out within the Generalized Gradient Approximation (GGA) for the exchange correlation potentials, using the Projector Augmented Waves (PAW) method[51-52], and a plane-wave basis set as implemented in the Vienna Ab-initio Simulation Package (VASP).[53-54] Van der Waals interactions were included at the DFT-D2 level using the second version of Grimme's dispersion corrections[55] combined with the revised Perdew-Burke-Ernzerhof (RPBE) functional[56] (RPBE-G06), which has been demonstrated to accurately describe molecule-metal interfaces[57]. A Tersoff-Hamman approximation was used within VASP to simulate STM images.[58]

ASSOCIATED CONTENT

**Supporting Information.** Include: density functional theory (DFT) methods; DFT-simulated STM image of a single TPPT terminated only on one end by a single Fe atom; DFT-calculated distances between atoms of TPPT-$Fe_3$-TPPT model; details on statistical count; and supporting evidence for the 3-Fe linkage structure between two TPPT molecules.

AUTHOR INFORMATION

Corresponding Author

* saburke@phas.ubc.ca, +1 (604)-822-8796 (Experiments)

* wji@ruc.edu.cn, +86-10-62515597 (Theory)19

Author Contributions

‡ A.S. and M.C. contributed equally.

Funding Sources

S. Burke, M. Capsoni, A. Schiffrin, T. Roussy, K. Cochrane, and A. Shaw acknowledge support from the Natural Sciences and Engineering Research Council, Canadian Foundation for Innovation, British Columbia Knowledge Development Fund, the Max Planck-UBC Centre for Quantum Materials (A.S.), the Canada Research Chairs Program (S.B.), the ARC Centre of Excellence for Future Low-Energy Electronics Technologies (A.S.) and the University of British Columbia. W. Ji and C.-G. Wang were supported by the National Natural Science Foundation of China (NSFC) under Grant Nos. 11274380 and 91433103, the Ministry of Science and Technology (MOST) of China under Grant No. 2012CB932704, and the Basic Research Funds of Renmin University of China from the Central Government under Grant No. 15XNH068. The Calculations were performed at the Physics Laboratory for High-Performance Computing of Renmin University of China and at the Shanghai Supercomputer Center.

Notes

Any additional relevant notes should be placed here.

ACKNOWLEDGMENTS

S. Burke, M. Capsoni, A. Schiffrin, T. Roussy, K. Cochrane, and A. Shaw acknowledge support from the Natural Sciences and Engineering Research Council, Canadian Foundation for Innovation, British Columbia Knowledge Development Fund, the Max Planck-UBC Centre for Quantum Materials (A.S.), the Canada Research Chairs Program (S.B.), the ARC Centre of Excellence for Future Low-Energy Electronics Technologies (A.S.) and the University of British

Columbia. W. Ji and C.-G. Wang were supported by the National Natural Science Foundation of China (NSFC) under Grant Nos. 11274380 and 91433103, the Ministry of Science and Technology (MOST) of China under Grant No. 2012CB932704, and the Basic Research Funds of Renmin University of China from the Central Government under Grant No. 15XNH068. The Calculations were performed at the Physics Laboratory for High-Performance Computing of Renmin University of China and at the Shanghai Supercomputer Center.

ABBREVIATIONS

MLCT, metal-to-ligand charge transfer; tpy, terpyridine; STM, scanning tunneling microscopy; STS, scanning tunneling spectroscopy, DFT, density functional theory; UHV, ultra-high vacuum; RT, room temperature; TPPT, terpyridine-phenyl-phenyl-terpyridine; pyr, pyridine; LDOS, local density of states; GGA, generalized gradient approximation; PAW, projector augmented waves; VASP, Vienna ab-initio simulation package; RPBE, revised Perdew-Burke-Ernzerhof; VIS/NIR, visible/near-infrared.

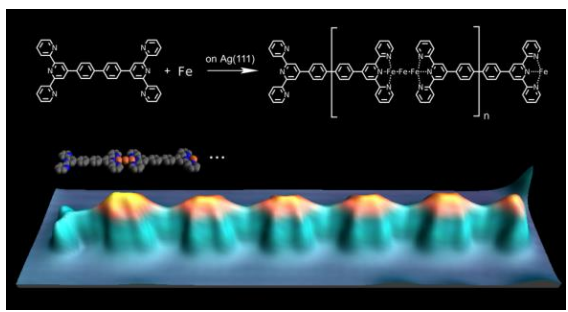

# Supplementary information

# Designing optoelectronic properties by on-surface synthesis: formation and electronic structure of an iron- terpyridine macromolecular complex

*Agustin Schiffrin*[||,~,†,§,‡], *Martina Capsoni*[||,‡], *Gelareh Farahi*[||], *Chen-Guang Wang*[^], *Cornelius Krull*[†], *Marina Castelli*[†], *Tanya S. Roussy*[||], *Katherine A. Cochrane*[#], *Yuefeng Yin*[§,±,†], *Nikhil Medhekar*[§,±], *Adam Q. Shaw*[||], *Wei Ji*[^,\*] *and Sarah A. Burke*[||,#,~,\*]

[||]Department of Physics and Astronomy, University of British Columbia, Vancouver British Columbia, Canada, V6T 1Z1

[~]Quantum Matter Institute, University of British Columbia, Vancouver British Columbia, Canada, V6T 1Z4

[†]School of Physics & Astronomy, Monash University, Clayton, Victoria 3800, Australia

[§]ARC Centre of Excellence in Future Low-Energy Electronics Technologies, Monash University, Clayton, Victoria 3800, Australia

[#]Department of Chemistry, University of British Columbia, Vancouver British Columbia, Canada, V6T 1Z1

[^]Department of Physics and Beijing Key Laboratory of Optoelectronic Functional Materials & Micro-nano Devices, Renmin University of China, Beijing 100872, People's Republic of China

[±]Department of Materials Science and Engineering, Monash University, Clayton, Victoria 3800, Australia

‡ These authors contributed equally.

\* saburke@phas.ubc.ca (Experiments)

\* wji@ruc.edu.cn (Theory)

## S1. Density Functional Theory (DFT) methods

Theoretical simulations were carried out using the General Gradient Approximation (GGA) for the exchange-correlation potentials[1], the Projector Augmented Waves (PAW) method,[2-3] and a plane-wave basis set as implemented in the Vienna Ab-initio Simulation Package (VASP).[4-5] Revised Perdew-Burke-Ernzerhof (RPBE),[6] with the second version of Grimme's dispersion corrections,[7] was adopted throughout all calculations. For the method above, van der Waals interactions were considered by the DFT-D2 level, which is known to give a better description of geometries and corresponding energies than those from the standard DFT.[8] The kinetic energy cutoff for the plane wave basis was set to 400 eV in configuration optimizations and increased to 600 eV for energy calculations. A supercell (8×7) consisting of 224 Ag atoms in 4 layers with at least 15 Å vacuum region was employed to model the chain configuration on the Ag(111) surface and an (11×7) one consisting of 308 Ag atoms in 4-layers for isolated molecules. The surface's Brillouin zones were sampled using the gamma point only in geometry optimizations and (3×3×1) in energy calculations. In geometry optimizations, all atoms except those at the bottom two Ag layers were fully relaxed until the residual force per atom was less than 0.02 eV/Å. Tersoff-Hamann approximations[9] were employed to execute scanning tunneling microscopy (STM) simulations in VASP. All simulated STM figures used consistent biases to experimental settings.

## S2. DFT-simulated STM image of a singly Fe-metalated TPPT molecule

Figure S1a-c show the DFT-optimized model, the DFT-simulated STM image, and the experimental STM image, respectively, of a single TPPT molecule with its right-hand terpyridine (tpy) group coordinated to a single Fe atom on Ag(111). The DFT-simulated image (Figure S1b) is in good agreement with the experimental STM data (Figure S1c), supporting the identification of the bright protrusion visible at the right-hand side of the molecule in Figure S1c as the result of the coordination of the tpy group with a single Fe atom. The left-hand side of the molecule shows the distinct "v-shape" characteristic of a non-Fe-terminated tpy group.

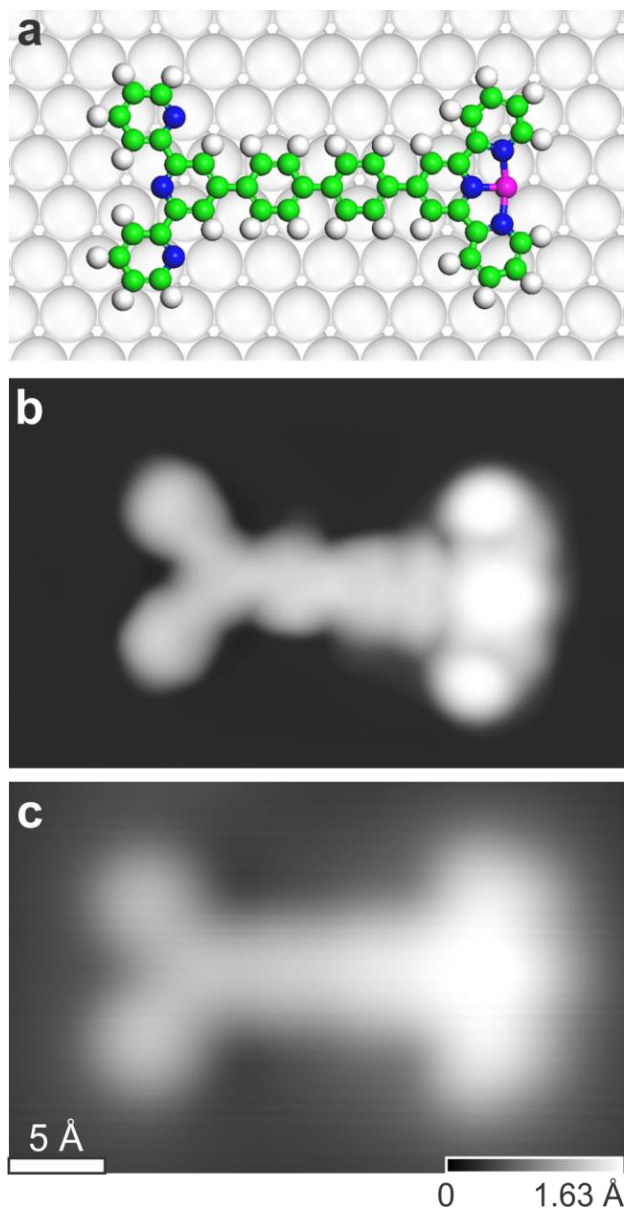

**Figure S1.** Single TPPT molecule with its right tpy group coordinated to a single Fe atom. (a) DFT-optimized model of TPPT-Fe on Ag(111) (green: carbon; white: hydrogen; blue: nitrogen; pink: iron). (b) Corresponding DFT-simulated STM image. (c) Experimental high-resolution STM image ($V_b$ = -200 mV, $I_t$ = 1 nA).

## S3. DFT-calculated distances between Fe atoms in the TPPT-Fe$_3$-TPPT complex

The N-to-Fe and Fe-to-Fe distances were calculated from the optimized DFT configuration and are summarized in Table S1 [the labels in Table S1 refer to the ones in Figure S2].

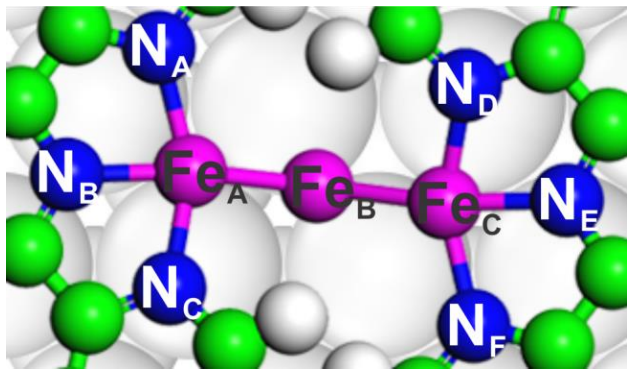

**Figure S2.** Top view of the DFT-optimized coordination center of the TPPT-Fe$_3$-TPPT structure. Nitrogen (N) and iron (Fe) atoms are labeled.

**Table S1.** DFT-calculated distances between N–Fe and Fe–Fe atoms for the TPPT-Fe$_3$-TPPT model of an infinite chain aligned along [1 -1 0] direction. The atoms labels correspond to the ones in Figure S2.

|  | Distance (Å) |
|:---:|:---:|
| Fe$_A$ – N$_A$ | 2.03 |
| Fe$_A$ – N$_B$ | 1.98 |
| Fe$_A$ – N$_C$ | 1.97 |
| Fe$_A$ – Fe$_B$ | 2.18 |
| Fe$_A$ – Fe$_C$ | 4.08 |
| Fe$_C$ – N$_D$ | 1.97 |
| Fe$_C$ – N$_E$ | 1.98 |
| Fe$_C$ – N$_F$ | 2.02 |
| Fe$_C$ – Fe$_B$ | 2.19 |

## S4. Supporting evidence for the 3-Fe linkage structure

Experimental STM data and DFT calculations hint towards a metal-organic coordination motif which consists of a linear tri-iron structure between two TPPT molecules. It is important to note that our STM topographic data do not allow to directly resolve the structure of such Fe coordination node at the level of single Fe atoms. Further studies are required for such direct experimental demonstration. Therefore, to provide further evidence that no other species (e.g. residual gas molecules, atoms from the substrate) participate in the metal-ligand coordination and in the chaining process, we have considered other alternative, plausible coordination motifs, other than TPPT-$Fe_3$-TPPT. Below, we assume that the linkage structure could hypothetically include water ($H_2O$) or carbon monoxide (CO) – molecules as the most likely bridging ligands available in ultrahigh vacuum (UHV) – as well as silver (Ag) atoms from the substrate, and provide experimental and theoretical evidence against these alternative.

    *a) Linkage involving water ($H_2O$) molecules*

Water molecules present in low concentrations in the UHV environment could be involved in the linkage between two TPPTs as a bridging ligand between Fe atoms.[10] However, the density of water molecules on the sample surface before iron deposition [$(0.78 \pm 0.32) \times 10^{-2}$ molecules/nm$^2$] is significantly less than the density of the links [$(2.27 \pm 0.41) \times 10^{-2}$ links/nm$^2$ after room temperature deposition and $(3.82 \pm 0.66) \times 10^{-2}$ links/nm$^2$ after annealing at 373 K] indicating that there is, at most, less than one $H_2O$ molecule per linkage, whereas a hydroxyl model would require at least one $H_2O$ molecule per Fe-Fe bond. We expect that linkages with such $H_2O$ bridging ligand would be imaged differently in STM than those without due to a substantially different electronic structure. As all the coordination linkages are imaged similarly, we do not see evidence for an $H_2O$ mediated structure even in small proportions, especially given the good agreement between our experiments and the DFT-calculated images providing support to the $Fe_3$ linkage model.

    *b) Linkage involving carbon monoxide (CO) molecules*

Molecular complexes where CO acts as a bridging ligand between Fe atoms, e.g., tri-iron dodecacarbonyl and related compounds, have been reported in literature.[11-12] However, the density of CO on the surface before the iron deposition is negligibly small (less than $H_2O$), leading us to eliminate any possible coordination model involving CO, which would require six CO per linkage to create a fully bridged structure.

   c) *Linkage involving silver (Ag) adatoms from the Ag(111) substrate*

Another possible alternative to the Fe-Fe-Fe linkage model is Fe-Ag-Fe (Figure S4), where a mobile Ag adatom from the Ag(111) substrate is incorporated in the coordination node. However, we have strong experimental evidence and supporting DFT calculations that contradict this possible structure. For instance, we performed annealing experiments on the TPPT/Ag(111) system, without the presence of Fe. After annealing at ~520 K – that is, at temperatures higher than those used for the Fe chaining – we observed the formation of small molecular chains with a kinked coordination node [Figure S3], distinctly different from the linear morphology obtained with Fe since. No such structures were observed at room temperature (RT). We hence conclude that the formation of such kinked structures is mediated by Ag adatoms diffusing off the substrate (i.e., step edges) at sufficiently high temperature, and that at lower temperatures (e.g., RT) the coordination centres do not include Ag and do not consist of mixed Ag-Fe cluster; the coordination linkages observed after RT deposition of Fe only contain Fe.

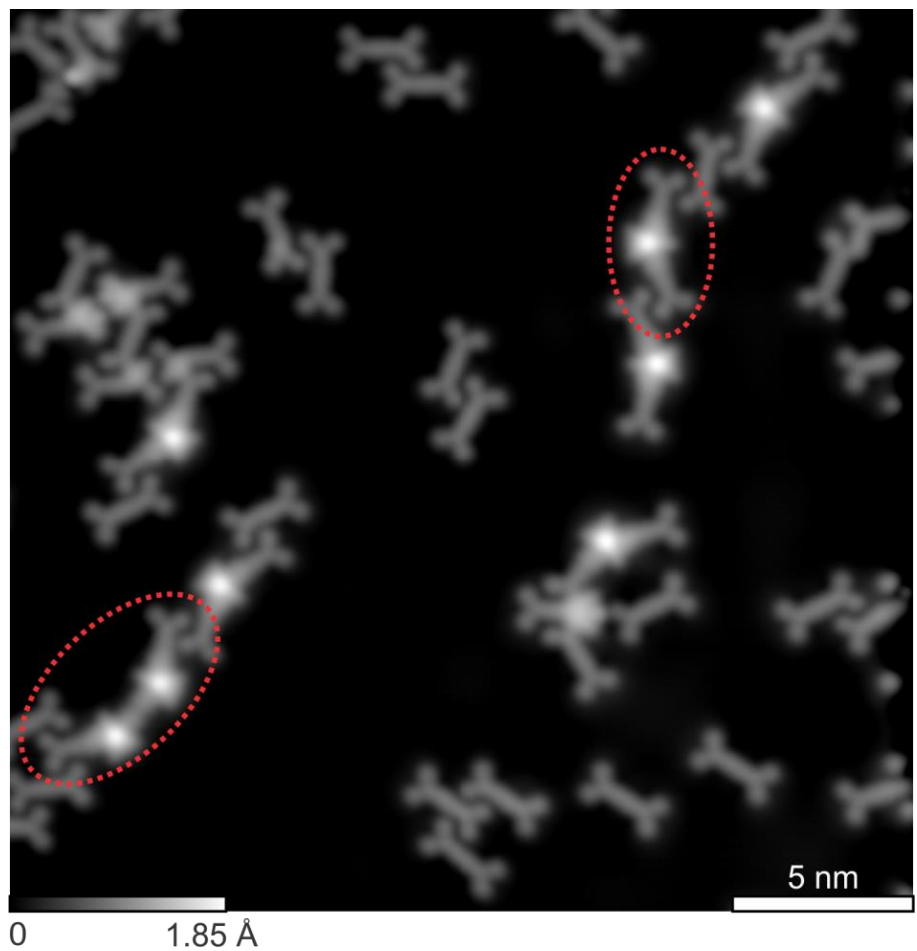

**Figure S3.** Constant-current STM image showing kinked chains (dashed red circled features) after annealing the TPPT/Ag(111) system at ~520 K for 30 minutes.

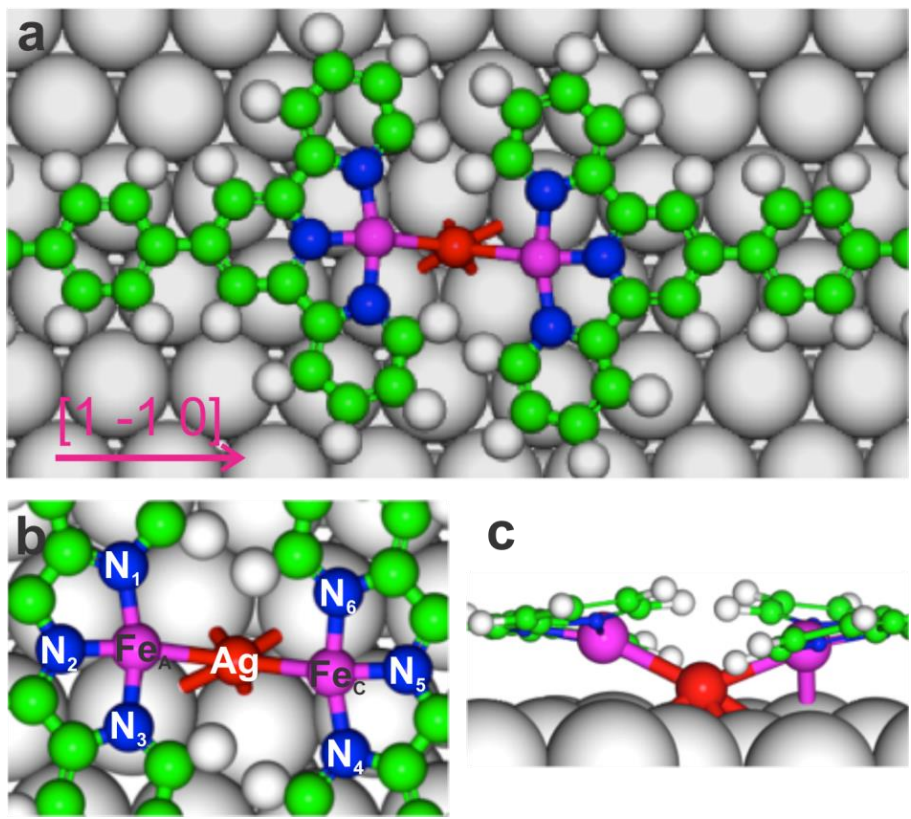

**Figure S4**. DFT simulated structure for an infinite chain of TPPT molecules linked by a Fe-Ag-Fe structure along the [1 -1 0] direction (a). Panel (b) is the top view of the chain, while (c) is the side view of Fe-Ag-Fe center.

A different scenario could occur after annealing the Fe/TPPT/Ag(111) system at 373 K. However, all of the linear metal-organic nanostructures are imaged identically before and after annealing, indicating that newly formed and extended chains are electronically and structurally equivalent. Additionally, no chains exhibiting the kinked morphology of the Ag-mediated linkages (Figure S3) were observed, further supporting the fact that at temperatures below 373 K there are no diffusing Ag adatoms that could contribute to chain formation.

As discussed in the main text, STM images at negative (positive) bias voltages show a protrusion (depression, respectively) at the Fe-Fe-Fe coordination centre. This is supported by DFT calculations (Figure 4 in main text). However, DFT simulations of the Fe-Ag-Fe link show that both negative (Figure S5a) and positive (Figure S5b) bias constant-current images exhibit two clear bright protrusions at the coordination site, and that their subtraction (Figure S5c) does not show the two very striking protrusions seen for the tri-iron case. Therefore, the imaging characteristics for a hypothetical Fe-Ag-Fe mixed linkage is not consistent with our experiments (see main text).

In particular, the DFT calculated adsorption energy for the 3-Fe configuration is found to be 0.7 eV lower than the Fe-Ag-Fe one, resulting in the former model to be more favorable.

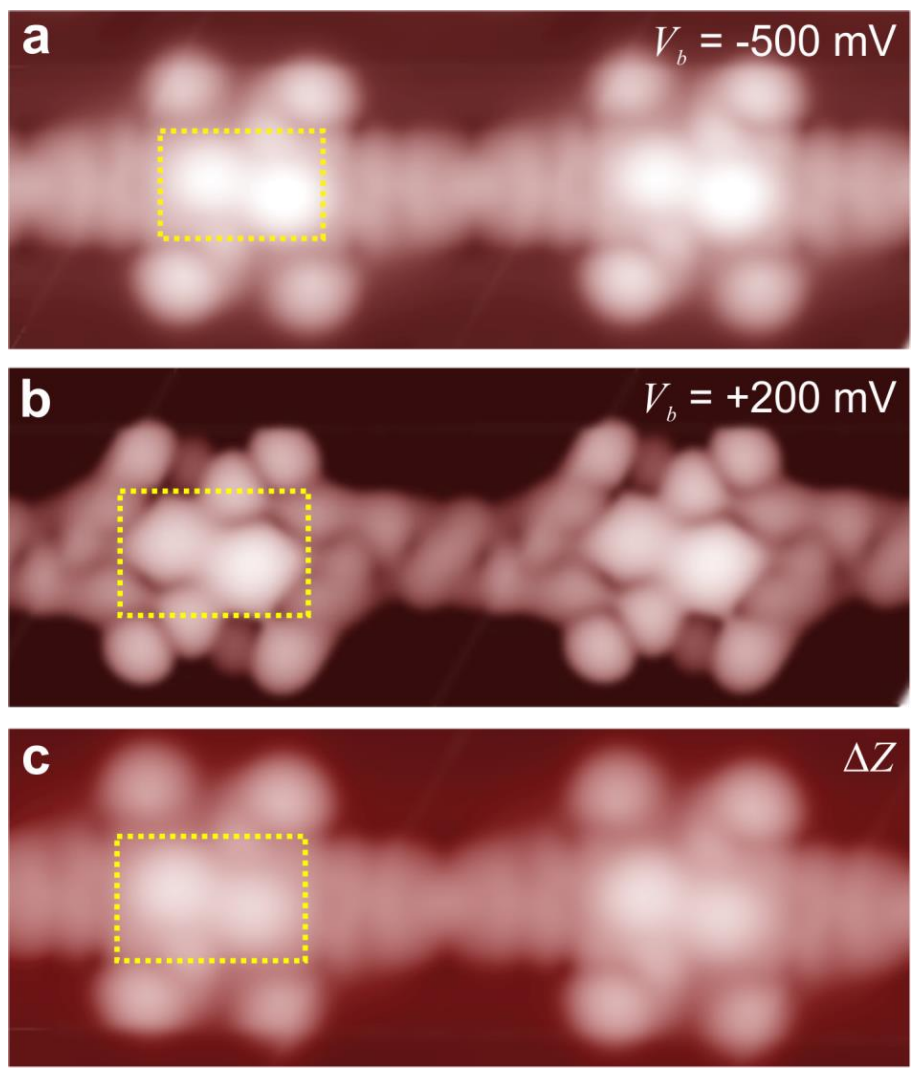

**Figure S5.** DFT-simulated constant-current STM images of a chain of TPPT molecules linked by the Fe-Ag-Fe structure shown in Figure S4 at -500 mV (a), +200 mV (b) and their subtraction (c) = (a) − (b).

**Table S2.** DFT calculated distances between atoms for the Fe-Ag-Fe (Figure S3) and the Fe-Fe-Fe structure (Figure S2). Labels refer to Figures S3b and S2.

|          | Distance              | $N_A$  | $N_B$  | $N_C$  | $Fe_B$ | $Fe_c$ | Ag   |
|----------|-----------------------|--------|--------|--------|--------|--------|------|
| Fe-Ag-Fe | $Fe_A$ (Å)            | 1.94   | 1.81   | 1.94   | -      | 4.68   | 2.55 |
| Fe-Fe-Fe | $Fe_A$ (Å)            | 2.03   | 1.98   | 1.97   | 2.18   | 4.08   | -    |
|          | $\Delta D$ (Å)        | -0.09  | -0.17  | -0.03  | -      | 0.6    | -    |

## S5. d$I$/d$V$ scanning tunneling spectroscopy

In the main text (Figures 3 and 5), we show (d$I$/d$V$)/($I$/$V$) scanning tunneling spectra for the single pristine TPPT molecule, singly metalated Fe-TPPT molecule, and Fe-TPPT nanochain. These STS data provide evidence that: (i) the chaining process is accompanied by a ~+0.3eV energy shift of ligand-related unoccupied states, indicative of a Fe-to-TPPT electron transfer; (ii) the highest (lowest) occupied (unoccupied) orbitals of the complex are dominated by Fe (ligand, respectively). Figure S6 below shows that these conclusions are maintained regardless of how the STS data are treated, that is, regardless whether (d$I$/d$V$), (d$I$/d$V$)/($I$/$V$) or retrieved local density of states (LDOS; according to Passoni el. in Ref. [13]) are considered. In the latter, the LDOS function was obtained by $\text{LDOS}(V_b) = (dI/dV)/T_{\text{asym}}$, where $V_b$ is the bias voltage and where the asymmetric transmission function $T_{\text{asym}}(V_b)$ was found by fitting d$I$/d$V$ on bare Ag(111) with $\text{LDOS}_{\text{Ag(111)}}(V_b) \cdot T_{\text{asym}}(V_b)$, where the local density of states $\text{LDOS}_{\text{Ag(111)}}$ of Ag(111) is assumed to be an error function (see caption of Figure S6).

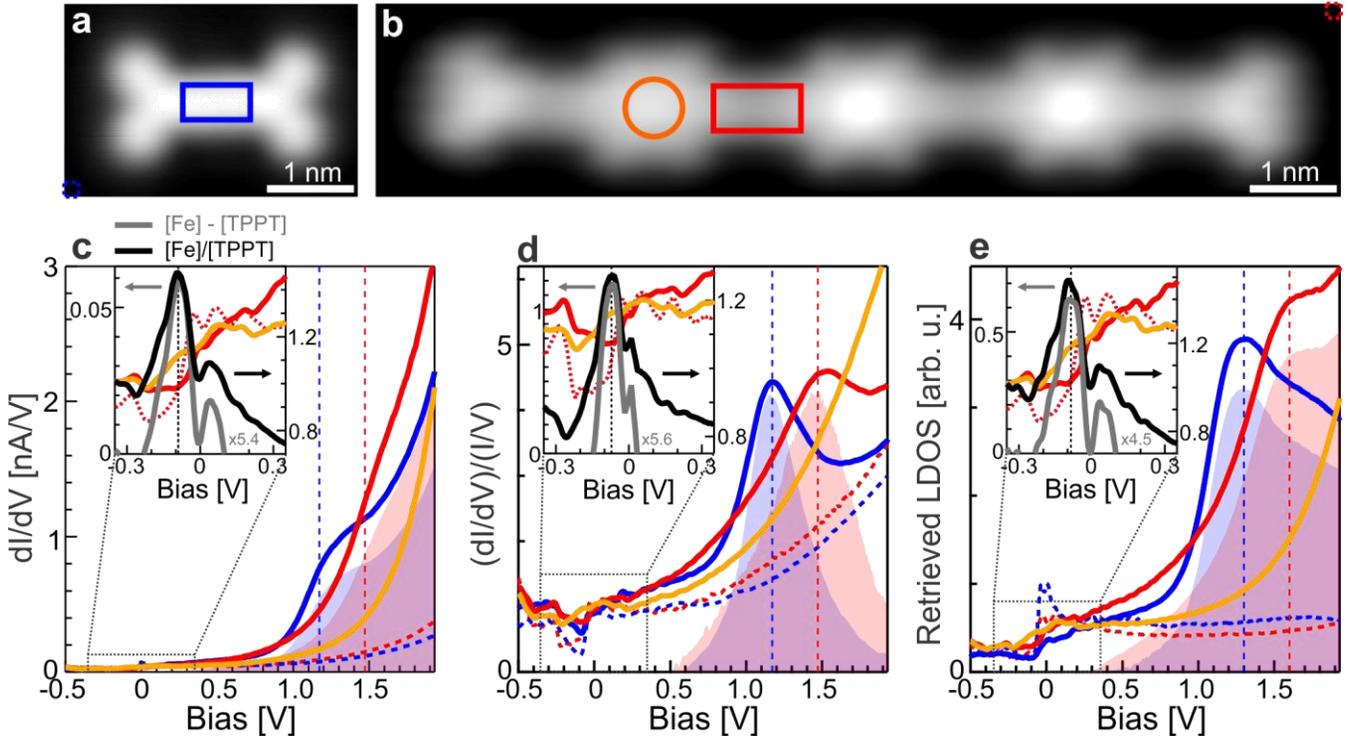

**Figure S6**. (a) STM images of a single, pristine TPPT ($V_b = -2.5$ V, $I_t = 1$ nA) and (b) of a Fe-TPPT nanochain ($V_b = -2.5$ V, $I_t = 2.5$ nA) on Ag(111). Rectangles and circle indicate areas where spectra (c) – (e) were acquired. (c) Average d$I$/d$V$, (d) (d$I$/d$V$)/($I$/$V$) and (e) retrieved local density of states (LDOS) spectra taken at centre of isolated, pristine TPPT molecule (blue), centre of TPPT in Fe-TPPT chain (red), and Fe coordination node (set point: $V_b = -2.5$ V, $I_t = 1$ nA for blue curves; $V_b = -2.5$ V, $I_t = 2.5$ nA for red and orange). The LDOS in (e) was retrieved according to Ref. [13]. In this approach, the $I(V)$ curve taken on bare Ag(111) was fitted with the one-dimensional Wentzel-Kramers-Brillouin expression of the tunneling current, with the LDOS of Ag(111) modelled by a constant added to an error function centered at the onset bias voltage $V_0$ of the Shockley surface state; $\text{LDOS}_{\text{Ag(111)}}(V_b) = a + b \cdot erf(c \cdot (V_b - V_0))$, with $a$, $b$, $c$, $V_0$ [as well as the work function and the tip-sample distance in the tunneling transmission function $T(V_b, E)$] as fitting parameters. The numerical derivative d$I$/d$V$ of $I(V)$ on bare Ag(111) was then fitted with the function $\text{LDOS}_{\text{Ag(111)}}(V_b) \cdot T_{\text{asym}}(V_b)$ where $T_{\text{asym}}(V_b) = A \cdot T(V_b, 0) + B \cdot T(V_b, eV_b)$, $A$, $B$ as fitting parameters. The retrieved LDOS was then calculated on the area of interest as $\text{LDOS}(V_b) = (dI/dV)/T_{\text{asym}}$. Dashed curves: spectra on bare Ag(111). Background filled curves (rescaled and offset for clarity): difference spectra resulting from the substraction of solid blue, orange and red curves with

S11

dashed curves. d*I*/d*V* spectra in (c) do not show clear peaks related to unoccupied molecular orbitals due to the strong exponential transmission function background. It is important to note that the (d*I*/d*V*)/(*I*/*V*) and retrieved LDOS curves in (d) and (e) minimize the influence of the exponential transmission and show clear peaks related to unoccupied states of the TPPT molecule. With respect to the single pristine TPPT, these peaks are shifted by ~+0.3 V for the molecule in Fe-TPPT nanochain. Insets: detail of spectra near the Fermi level. Solid black (grey) curves correspond to the ratio (difference, respectively) between spectra taken on the Fe node (orange) and ligand (red) in the Fe-TPPT nanochain. Grey difference curves were rescaled for clarity. Both ratio (black) and difference (grey) curves show a clear feature near the Fermi level, providing strong evidence that the highest occupied electronic levels of the Fe-TPPT system are dominated by Fe states.

## S6. Bias-dependent STM of single pristine and doubly metalated TPPT

In Figure 3 of the main text, we show bias-dependent STM imaging of a single, doubly metalated Fe-TPPT-Fe molecule, as well as related (d*I*/d*V*)/(*I*/*V*) STS. The corresponding substracted STM topograph (Figure 3c of main text) shows a protrusion at the centre of the metalated tpy group, that we associate with the location of the Fe atom. Figure S5 below corroborates this deduction, showing that a pristine, non-metalated tpy does not exhibit such protrusion.

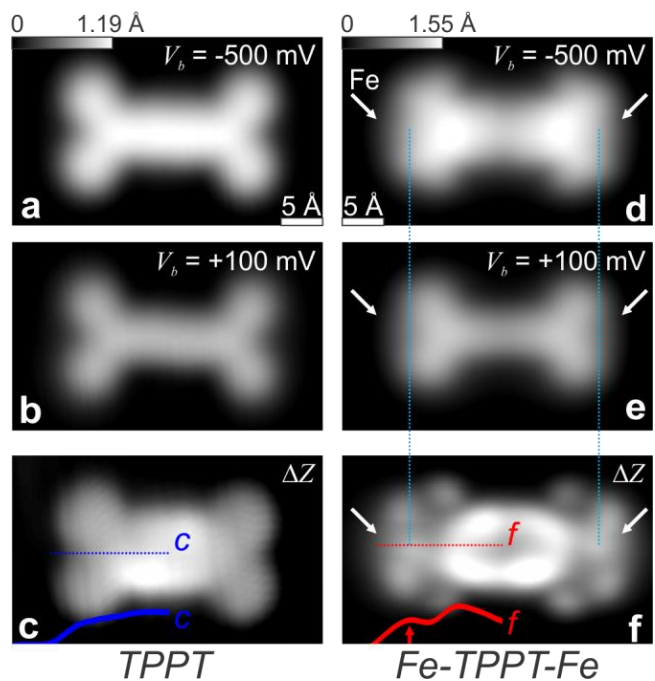

**Figure S7.** (a, d) Negative (-500 mV) and (b, e) positive (+100 mV) bias STM images of single, pristine TPPT molecule (a, b; $I_t$ =200 pA) and of TPPT molecule with both tpy groups metalated with Fe adatoms (d, e; $I_t$ =200 pA). (c) Difference STM topographic map resulting from subtraction of (a) with (b). Solid blue curve: height profile along blue dashed line *c*. (f) Same as (c), but resulting from subtraction of (d) with (e). White arrows and vertical cyan dashed lines indicate the position of single Fe adatoms. Solid red curve: height profile along red dashed line *f*. Red arrow indicates a protrusion related to the Fe adatom.

## S7. Spectroscopic signature of isolated Fe and Fe in nanochain

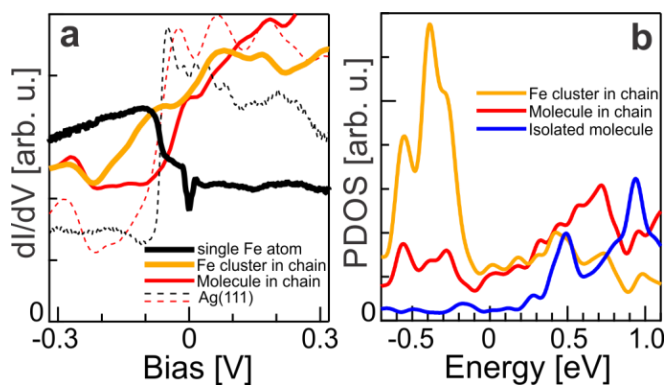

**Figure S8.** (a) d$I$/d$V$ spectra acquired on isolated Fe adatom on Ag(111) (black; set point: $V_b$ = -0.4 V, $I_t$ = 2.5 nA), at Fe coordination node in Fe-TPPT chain (orange; set point: $V_b$ = -2.5 V, $I_t$ = 2.5 nA) and at

S13

centre of TPPT molecule in Fe-TPPT chain (red; set point: $V_b$ = -2.5 V, $I_t$ = 2.5 nA). Spectroscopic signature[14] of isolated Fe adatom is due to localization of the Ag(111) Shockley surface state (peak at $V_b$ ~ -100mV) and Kondo resonance (dip at $V_b$ = 0). This signature is qualitatively different from the Fe cluster in the metal-organic nanochain, where a feature is observed at larger bias voltage, and no Kondo resonance is visible. (b) Projected density of states of the metal-organic nanochain, derived from DFT calculations (orange: contribution from Fe states; red: from TPPT molecule in nanochain; blue: isolated TPPT). Electronic states of the Fe coordination node contribute dominantly to the occupied density of states of the system near the Fermi level, as seen in our experiments.